\documentclass[twocolumn,showpacs,preprintnumbers,amsmath,amssymb]{revtex4}

\usepackage{graphicx}
\usepackage{bm}

\begin{document}
\title{Wave Energy Amplification in a Metamaterial based Traveling Wave Structure}
\author{Y. S. Tan and R. Seviour}
\affiliation{Dept. Physics, \\ Lancaster University,\\LA1 4YB UK.}
\date{\today}

\begin{abstract}
We consider the interaction between a particle beam and a propagating electromagnetic wave in the presence of a metamaterial. 
We show that the introduction of a metamaterial 
gives rise to a novel dispersion curve which determines a unique wave particle relationship, via the frequency dependence of the metamaterial and the novel ability of metamaterials to exhibit simultaneous negative permittivity and permeability. Using a modified form of Madey's theorem we find that the novel dispersion of the metamaterial leads to a amplification of the EM wave power.
\end{abstract}


\maketitle
\section{Introduction}
 Metamaterials are artificial macroscopic composites with a periodic cellular structure which produce two or more responses
not available in nature
in response to a specific excitation \cite{walser}. 
In this paper we focus on a class of metamaterials with negative permittivity and permeability, termed "Double NeGative materials" (DNG). Veselago \cite{ves} showed that a DNG material can 
control the phase of the EM field
to give an effective negative index of refraction, achieved by the DNG  presenting a relatively
high opposing EM field.
 Realization of DNG materials has been achieved using two different techniques;  a lattice of split-ring resonators and thin wires \cite{3},
and  loaded transmission lines \cite{4,5,6,7}. These metamaterials have been used to construct a range of novel microwave devices such as antennas \cite{6,7,8}, phase-shifters \cite{9,10}, couplers \cite{12,13}, broadband/compact power-dividers \cite{15} and other devices such as beam steerers, modulators, band-pass filters and lenses.

\begin{figure}
\includegraphics[width=3in]{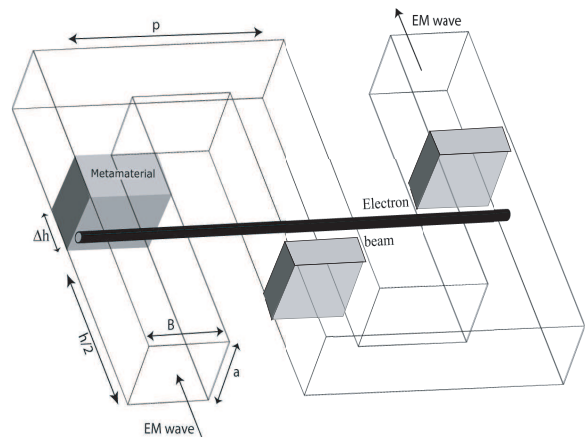}
\caption{The traveling wave structure \textbf{considered} here, consisting of a folded waveguide with a metamaterial insert, the electron beam passes through the middle of the structure.}
\label{fig0}
\end{figure}

In this paper we consider the application of metamaterials to
 Traveling Wave Tubes (TWT).
The TWT  proposed in the 1940's by Kompfner \cite{twtpap} remains the  driving technology for many  applications ranging from communications to radar. The principle of the TWT is to
amplify an applied EM wave of a specific frequency. This is achieved by passing the EM wave through a Slow
Wave Structure (SWS) simultaneously with an electron beam, such that wave and beam pass through the structure with similar velocities, for the EM wave this is determined by the dispersion relationship of the SWS .
The interaction between electron beam and EM field results in an energy transfer from beam to wave. 
To date three papers \cite{nasa1, nasa2, nasa3} have considered metamaterials in TWTs, all used  metamaterials to line the side of the structure to minimise losses and increase efficiency. 
We consider the case where the metamaterial forms part of the SWS. The basis of our structure is a Folded Waveguide (FWTWT) with a metamaterial inset, as shown in figure \ref{fig0}, where a $TE_{01}$ wave propagates along
the waveguide.  
We
introduce metamaterial at the interaction region between beam and wave, controlling the FWTWT dispersion relationship via the metamaterial,
to define a unique beam-wave interaction, triggering a novel gain-frequency phenomena.


\section{methodology}
For effective energy transfer between beam and EM wave
the phase velocity (determined by the dispersion) of the wave must approximately match
the velocity of the electron beam. In the conventional FWTWT this is achieved via the periodicity of the folded waveguide \cite{slow} to slow down the wave, generating Spatial Harmonics Wave Components (SHWC)
parallel to the beam. The SHWC interact with the beam resulting in energy transfer. By a superposition of the spatial harmonics $(\vec{E}_m(x,y))$ the field parallel to the beam can be expressed by
 Floquet's theorem  \cite{har} as;

\begin{equation}
\vec{E}(z)=\sum_{m=-\infty}^{\infty} \vec{E}_m(x,y) e^{-i\beta_m z},  \beta_m=\beta+\frac{2m\pi}{p}
\label{harmo}
\end{equation}
 To ensure that the phase of the EM field is the same at each point where wave and beam interact, the wave takes the long path around the folded wave guide, hence
 the period in the beam frame of reference is half the geometrical period of the structure shown in figure1. This phase shift results in a propagation constant $\beta$ \cite{park}:

\begin{equation}
\beta= \beta_0 \left( 1+ \frac{h}{p}\right) + \frac{\pi}{p}= \omega t=\omega (p/v_e)\\
\label{harmo2}
\end{equation}

\begin{equation}
\beta_0 = c^{-1} \sqrt{\omega^2-\omega_c^2}
\label{harmo3}
\end{equation}

 $\omega$ is the frequency of the incident EM wave, $\omega_c$ is the waveguide cutoff frequency, $p$ and $h$ are the period and height of the structure, and $\beta_0$ is the $TE_{01}$ rectangular
waveguide propagation constant. This form of \textbf{equation} \ref{harmo2} ensures that wave arrives at the interaction region with the same phase as previously seen by the  beam. Using equations \ref{harmo} and \ref{harmo2} we can derive the dispersion relationship for the $m^{th}$ SHWC;

\begin{equation}
\omega=\sqrt{\omega_c^2+\frac{c^2}{\left( 1+h/p \right) ^2 }\left( \beta_m -\frac{2m\pi+\pi}{p}\right)^2}
\label{nordisp}
\end{equation}

 For the
 conventional TWT the dispersion relationship and hence the phase velocity ($\omega/\beta$) is solely defined through the physical dimensions of the structure. We now consider the effect caused by inserting a metamaterial, of length $\Delta h$ into the waveguide at the point of interaction between wave and beam.  In a macroscopic medium the interaction with an electromagnetic wave is described through the constitutive relationships;
\begin{equation}\label{con}
\begin{array}{c}
\hat{D}=\epsilon_0 \hat{E} + P = \epsilon \hat{E} \\
\hat{B}=\mu _0 \hat{H} + M = \mu \hat{H}
\end{array}
\end{equation}

where $\hat{D}$ and $\hat{B}$ are the averaged electric and magnetic flux density, $\hat{E}$ and $\hat{H}$ are the averaged electric and magnetic
field, P is the averaged polarization (electric dipole moment density), and M is the averaged magnetization
(magnetic dipole moment density). $\epsilon$ and $\mu$  the  permittivity and permeability of the material define how
the material responds to an applied EM field. Equation \ref{con} represents a slab
of a isotropic, homogeneous material and can be described by a dispersive
Drude (permittivity) and Lorentz (permeability)
model \cite{8}. Where the  permittivity and permeability can be defined as:

\begin{equation}\label{mp1}
\epsilon_r (\omega)=\epsilon_{\infty}-\frac{\epsilon_{\infty}\omega^2 _p}{\omega(\omega-iv_c)}  \\
\end{equation}

\begin{equation}\label{mp2}
\mu_r (\omega)=\mu_{\infty} +\frac{(\mu_{s}-\mu_{\infty})\omega^2 _0}{\omega^2 _0-\omega^2+i\omega \delta}
\end{equation}

\begin{figure}
\includegraphics[width=3in]{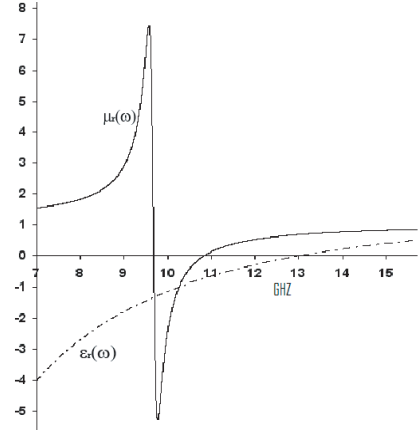}
\caption{The real component of equations \ref{mp1} and \ref{mp2}, with the parameters
$\epsilon_{\infty}=1.62$, $\mu_{\infty}=1.12$, $\mu_{s}=1.26$, $\omega_p=2 \pi 14.62$ GHZ, $\omega_0=2 \pi 9.56$ GHz, \textbf{$v_c=3.07 \cdot 10^7$}, and \textbf{$\delta=1.24 \cdot 10^9$}.}
\label{fig1}
\end{figure}

 $\epsilon_{\infty}$
 the permittivity in the high frequency limit, $\omega _p$ the radial plasma frequency, $v_c$ collision frequency. $\mu_{s}$ $(\mu_{\infty})$
 the permeability at the low(high) frequency limit, $\omega_0$ radial resonant frequency, and $\delta$ damping frequency. Considering equations \ref{mp1} \& \ref{mp2}, as shown in figure \ref{fig1}, we note that  $\epsilon_r$ and $\mu_r$  show a  frequency range
 over which both $\epsilon_r$ and $\mu_r$ are negative, in this region the material is a DNG. 
Assuming  the metamaterial is isotropic the phase constant ($\beta_{mm}$) describing  $TE_{01}$ propagation is no different to that describing a dielectric,
\begin{equation}\label{bmm}
\beta_{mm}(\omega)= c^{-1} \sqrt{\omega^2 \epsilon_r(\omega) \mu_r(\omega) - \omega_c^2}
\end{equation}

This results in the modification of equation \ref{harmo2} to accommodate the phase shift due to the metamaterial region as;
\begin{equation}
 \gamma_n = \beta_{0} \frac{p+h- \Delta h}{p}+ \beta_{mm}(\omega) \frac{\Delta h}{p}+(2n+1)\frac{\pi}{p}
\label{bn}
\end{equation}
 Rearranging equation \ref{bn} yields the dispersion relationship, equation \ref{wmm}. Comparing the dispersion relationships for the non-metamaterial  structure (equation \ref{nordisp}) and the metamaterial structure (equation \ref{wmm}), in the non-metamaterial the dispersion (and wave velocity) is defined by the dimensions of the waveguide, whereas in the metamaterial structure the dispersion now exhibits a frequency material dependence.
\begin{equation}
 \omega= \sqrt{ c^2  \left( \gamma_n -  \left(\frac{\pi(2n+1)}{p}+ \beta_{mm}(f) \frac{\Delta h}{p} \right) \right)^2 \alpha +\omega_c^2 }
\label{wmm}
\end{equation}
\begin{equation}
\alpha=\left(\frac{p}{p+h-\Delta h} \right)^2
\end{equation}

\begin{figure}
\includegraphics[width=3in]{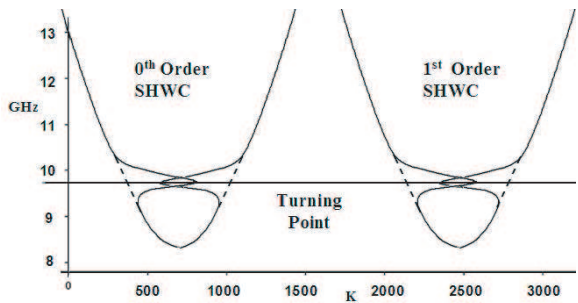}
\caption{Zeroth and first order SHWC dispersion curves, the solid line shows the dispersion for the metamaterial TWT and the doted lines shows the dispersion for the conventional TWT.}
\label{disp}
\end{figure}

Hence for a given form of $\epsilon_r(\omega)$ and  $\mu_r(\omega)$ we can define a unique dispersion curve.
The form of $\epsilon_r(\omega)$ and  $\mu_r(\omega)$ is specified by careful design of the metamaterial to give additional control over the dispersion. Considering a metamaterial with the following material properties,
    $\epsilon_\infty = 1.62$,
    $\mu_\infty= 1.12$,
    $\mu_s = 1.26$,
    $\omega_p =2 \pi 14.62 \cdot 10^9$,
    $\omega_0=2 \pi 9.56 \cdot 10^9$,
    $v_c =3.07 \cdot 10^7  $ m/s,
    $\delta=1.24 \cdot 10^6$,
    $\Delta h=5 $  mm, we can examine the form of the dispersion relation has on the SHWC, as shown in figure \ref{disp}. The dotted line in figure \ref{disp} shows the dispersion curve of a conventional FWTWT (equation \ref{nordisp}) and the solid curve is the metamaterial loaded FWTWT. The presence of the metamaterial has marked effect on the dispersion of the SHWC, in particular the turning point at 9.7 GHz in the dispersion.
     From figure \ref{fig1} we note the turning point in figure \ref{disp} relates to the start of the regime where the metamaterial has both $\epsilon_r,\mu_r<0$ (i.e. a DNG).

The dispersion curve of each harmonic can be considered as consisting of two branches, "left" and "right" of the phase constant associated with the cut-off frequency (the minima in $\omega$).  In conventional FWTWTs, the group velocity ($v_g=\partial \omega / \partial \beta $) which indicates the direction of power flow, in the left branch is $v_g < 0$, while in the right branch is $v_g>0$.
In the metamaterial FWTWT the dispersion curve exhibits a turning point giving rise to a discontinuity in both group velocity and phase velocity, at the turning point the wave does not have a coherent group. This feature in the dispersion curve creates a frequency regime where harmonics of both left and right branches can have $v_g>0$ and $v_g < 0$. Above the turning point individual harmonics consist of both left and right branches with positive group velocity, which implies both branches flow forward with EM power.  However analysis of the spatial harmonics verifies that only the right branch components are physically triggered with the forward EM wave.


To determine the energy transfer from beam to wave we utilize Madeys theorems \cite{mad}, which relates the phase averaged energy spread to the phase averaged energy change experienced by an electron as it passes through the interaction region. The theory uses the first two terms of a perturbation expansion of the Lorentz force based on
 a single particle analysis  in which a uniformly distributed electron beam, ignoring space charge, is injected into the system. The theorem links the spontaneous emission of photons by a single electron passing through the structure to the stimulated emission.
 The first perturbation energy term of the Lorentz equation $\gamma_1$, is taken at entry and exit from the structure, the difference $\Delta \gamma_1$ is averaged over the phase of the applied EM field to yield equation \ref{lor}. This change in energy $\Delta \gamma_1$ relates to the classical spontaneous power spectrum from Maxwell's equations from the beam\cite{mad}.

\begin{equation}
 <\Delta \gamma_1^2>=\left < \left( \int dr \frac{-e\vec{E} \cdot  v_0}{m_0 c^2} \right) ^2 \right>
\label{lor}
\end{equation}


The 2nd order perturbation term relating the energy change in the beam due to stimulated emission is  a consequence of a generalized framework in Hamiltonian mechanics and is given by;

 \begin{equation}
<\Delta \gamma_2>=  \frac{1}{2} \frac{d}{d \gamma}  <\Delta \gamma_1^2>
\label{lor2nd}
\end{equation}

If we now consider the electron beam as consisting of $N$ electrons entering the system every second we can then write the power increase in the EM wave as;

\begin{equation}
\Delta P=  -\frac{1}{2} \frac{d}{d \gamma}  <\Delta \gamma_1^2> m_0 c^2 N
\label{P}
\end{equation}

The complexity of the full form of $\Delta P$ requires a numerical approach to calculation, a full derivation will follow in a future publication. With a change of variable and via substitution of the above terms we can express the change in power as;
\begin{equation}
\Delta P=\frac{{\rm \omega }^{{\rm 2}} {\rm \mu }}{{\rm \beta }_{{\rm 0}} } \frac{L^3}{{2 ab}}
Z^{2}
 \frac{{\rm d}}{{\rm dX}} \left(\frac{{\rm sin}^{{\rm 2}} \left({\rm X}\right)}{{\rm X}^{{\rm 2}} } \right)\frac{{\rm c}}{{\rm \gamma }^{{\rm 3}} } \frac{{\rm 1}}{{\rm v}_{{\rm e}} ^{{\rm 3}} } \left({\rm mc}^{{\rm 2}} {\rm I_b /e}\right)
\label{P}
\end{equation}

$Z=\frac{{\rm e}}{{\rm mc}^{{\rm 2}} } \frac{{\rm b}}{{\rm p}} {\rm sinc}\left({\rm \beta }'_{{\rm n}} \frac{{\rm b}}{{\rm 2}} \right)$

${\rm \gamma }={\rm 1}+{{\rm V}_{{\rm acc}}  \mathord{\left/{\vphantom{{\rm V}_{{\rm acc}}  \left({\rm m}_{{\rm 0}} {\rm c}^{{\rm 2}} \right)}}\right.\kern-\nulldelimiterspace} \left({\rm m}_{{\rm 0}} {\rm c}^{{\rm 2}} \right)} $

${\rm X}=\left(\frac{{\rm \omega }}{{\rm v}_{{\rm e}} } -{\rm \beta }'_{{\rm n}} \right)\frac{{\rm L}}{{\rm 2}} $

\begin{figure}
\includegraphics[width=3in]{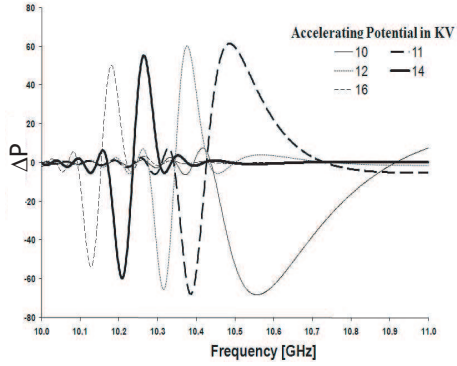}
\caption{Increase in EM power due to interaction with an electron beam, for different electron beam energies.}
 \label{dsde}
\end{figure}

This form  introduces the beam current \textbf{($I_{b}$)} and the voltage though which the beam has been accelerated ($V_{acc}$). The sign of $X$ is determined by the differences in velocity between wave and beam, which for $X<0$ indicates that initial the beam velocity is greater than the wave velocity. When the beam has higher velocity energy is exchanged from beam to wave until the velocities synchronism. Figure \ref{dsde} is a plot of equation \ref{P} for several different $V_{acc}$. Where maximum energy exchange occurs for $X=-1.3$. As expected by design the largest $\Delta P$ occurs for an accelerating voltage of \textbf{$11KV$} at $10.5$ GHz.

\section{Conclusion}

This paper demonstrates that  a metamaterial FWTWT offers an additional factor to controlling the gain-frequency characteristics, \textbf{compared to the conventional FWTWT where the characteristics depend on the waveguide dimensions.}
Figure 4, the change in power between wave and beam, shows that as the accelerating potential is increased  the frequency at which maximum energy exchange is achieved  is shifting towards lower frequency.
Although we note that even for large differences in accelerating voltage the frequency shift is small, this offers a precise way to tune the frequency of operation.

The disadvantages are  that the  design is  bandwidth limited, and highly dependent on the metamaterial used. Inherent ohmic losses associate with the MM are unavoidable. Future work in this area is to consider the use of a MM which offers a broadband of negative behavior, and a full discussion on the derivation of equation[15].

\end{document}